\documentclass[aps,groupedaddress]{revtex4}
\usepackage{epsfig}
\usepackage{graphicx}
\usepackage[T1]{fontenc}
\usepackage{ae}
\usepackage{ulem}
\usepackage{color}
\usepackage[latin1]{inputenc}
\usepackage{amssymb,amsbsy,amsmath}
\usepackage{bbm}

\newtheorem{assu}{Assumption}

\input ulem.sty

\begin{document}

%%%%%%%%%%%%%%%%%%%%%%%%%%%%%%%%%%%%%%%%%%%%%%%%%%%%%%%%%%%%%%%%%%%%%%%%%%%%%

\title[JTE]
      {Stochastic thermodynamics  of a finite quantum system coupled to two heat baths}
\author{Heinz-J\"urgen Schmidt$^1$ and Jochen Gemmer$^1$ }
\address{$^1$  Universit\"at Osnabr\"uck,
Fachbereich Physik,
 D - 49069 Osnabr\"uck, Germany}

%\tableofcontents

\begin{abstract}
We consider a situation where an $N$-level system (NLS) is coupled successively to two heat baths with different temperatures
without being necessarily thermalized and approaches a steady state.
For this situation we apply a general Jarzinski-type equation and conclude that heat and entropy is flowing from the
hot bath to the cold one.
The Clausius relation between increase of entropy and transfer of heat divided by a suitable temperature
assumes the form of two inequalities. Our approach is illustrated by an analytical example.
For the linear regime, i.~e., for small temperature differences between the two heat baths we derive
an expression for the heat conduction coefficient.
\end{abstract}

\maketitle
%%%%%%%%%%%%%%%%%%%%%%%%%%%%%%%%%%%%%%%%%%%%%%%%%%%%%%%%%%%%%%%%%%%%%%%%%%%%%%%%%%%%%%%%%%%%%%%%%%%%%%%%%%%%%%%%%%%%%%%%%%%%%%%
\section{Introduction}\label{sec:Intro}
%%%%%%%%%%%%%%%%%%%%%%%%%%%%%%%%%%%%%%%%%%%%%%%%%%%%%%%%%%%%%%%%%%%%%%%%%%%%%%%%%%%%%%%%%%%%%%%%%%%%%%%%%%%%%%%%%%%%%%%%%%%%%%%

The study of the non-equilibrium thermodynamics of heat conduction has a long history.
As an example consider the sequence of papers of Lebowitz and Spohn \cite{LS78,SL87,LS99} dealing with both
classical and quantum aspects of this problem.
A considerable number of papers concerned with thermal conduction for quantum systems
consider a kind of chain of small quantum systems connected to the two heat baths at their ends,
and focus on the questions whether a local thermal equilibrium is reached in the small quantum systems,
whether a constant temperature gradient exists and whether the heat transfer rate is proportional to this gradient
(``Fourier's law"), see, e.~g., \cite{STM96,MHGM03,S03,MGM04,MMG05,JHR06,WHBGM07,P11}.

The present paper is \textit{not} concerned with the validity of Fourier's law for special systems, but rather assumes a general $N$-level system (NLS)
with alternating contact with two heat baths and studies the heat and entropy transfer in the steady state assumed after some time,
without further analyzing the internal structure of the NLS.
For this purpose we use new methods that have been developed during the last decades,
in particular, the approach via fluctuation theorems, see, e.~g. \cite{LS99}.  The famous Jarzynski equation represents one of the rare exact results in nonequilibrium statistical mechanics. It is originally a statement of the expectation value of the exponential of the work $\left\langle e^{-\beta \,w}\right\rangle $
performed on a system initially in thermal equilibrium with
inverse temperature $\beta$, but may be far from equilibrium after the work process. This equation was first formulated for classical
systems \cite{BK77,BK81,J97} and subsequently proved for quantum systems \cite{P00,K00,T00,M03}.
Extensions for systems initially in local thermal equilibrium \cite{T00}, micro-canonical ensembles \cite{TMYH13}, and large-canonical ensembles
\cite{SS07,SU08,AGMT09,YTC11,CHT11,E12,YKT12} have been published.

The most common interpretation of the quantum Jarzynski equation is to consider sequential measurements, see \cite{P00}.
This approach is also followed in the present work. The general framework for such an approach was outlined in \cite{SG20a} and \cite{SG20b}.
It is neither quantum mechanical nor classical {\it per se} and will be referred to as ``stochastic thermodynamics'' in the present work.
We have recently applied this method to the problem of the interaction of an NLS with a single heat bath \cite{SSG21}.
Here we will extend this work to the interaction with two heat baths of possibly different temperature.
Again, it is possible to describe the influence of the heat baths by an $N\times N$ stochastic matrix $T$ that gives the conditional
probability of transitions between the $N$ levels of the NLS. Moreover, it is sensible to identify the steady state regime with
fixed points of $T$.

\begin{figure}[htp]
\centering
\includegraphics[width=0.5\linewidth]{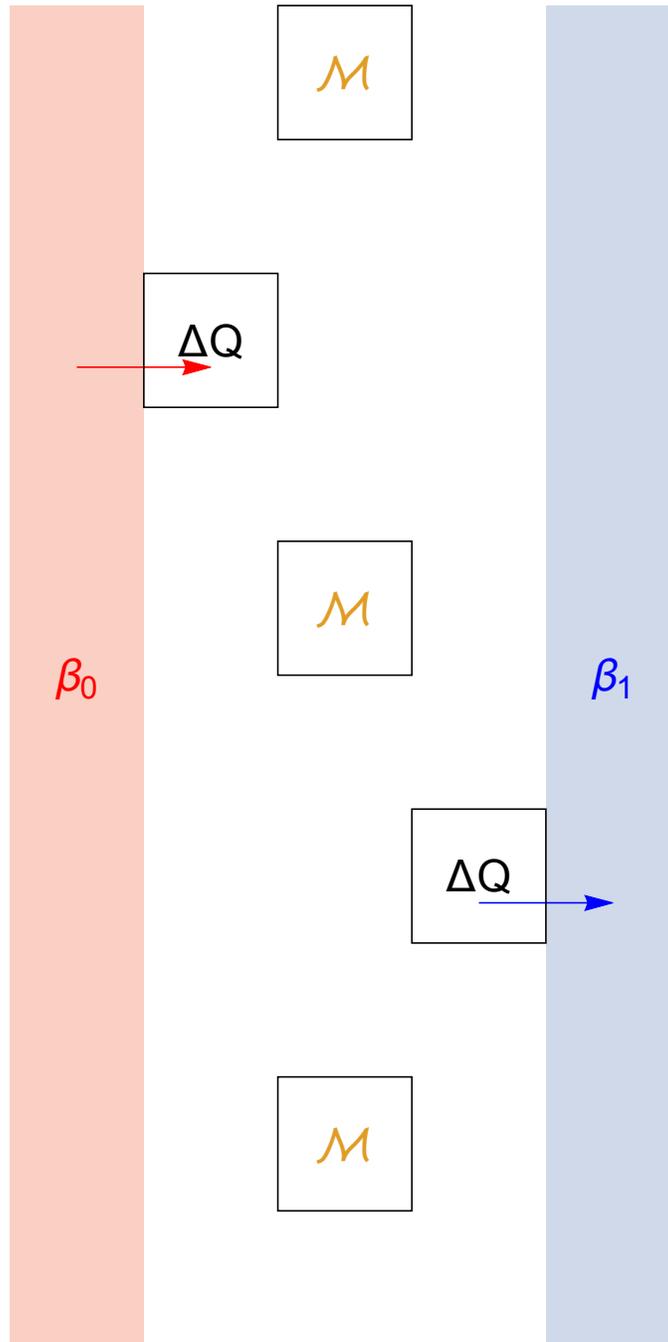}
\caption{Schematic sketch of one cycle of the alternate interaction of the NLS (black squares) with two heat baths of inverse
temperature $\beta_0<\beta_1$, resp., interspersed with energy measurements (yellow ${\mathcal M}$).
At steady state the heat $\Delta Q$ absorbed by the NLS during contact with the first heat bath
equals in amount the heat emitted during contact with the second heat bath.
}
\label{FIGTHB}
\end{figure}

However, it is not readily possible to infer the heat flow between the NLS and heat baths from the change in energy of the NLS, analogously
for the entropy flow. At steady state the corresponding changes of energy and entropy would vanish although the flow may be non-zero.
To cope with this problem we consider a step-wise interaction with the heat baths, see Figure \ref{FIGTHB}:
We start with a preparatory energy measurement and a contact only with the first heat bath, followed by a second energy measurement,
together described by a stochastic matrix $T^{(0)}$. This is followed by a separate contact with the second heat bath
and another measurement, together described by $T^{(1)}$. The total interaction plus measurements
is hence given by the product matrix $T=T^{(1)}\,T^{(0)}$. Moreover, we will assume, similarly as in \cite{SSG21}, that the
single stochastic matrices $T^{(i)}$ have invariant probability distributions of Gibbs type thereby introducing the two inverse
temperatures $\beta_i,\;i=0,1$ of the heat baths. This enables us to investigate the dependance between direction of heat and entropy flow
and the temperature difference, at least for the steady state.
The main message is that, although it would be very difficult to calculate the stochastic matrices $T^{(0)}$ and $T^{(1)}$ for ``real" systems,
the validity of the second law-like statement about the correct direction of heat and entropy flow
only depends on some general properties of these matrices and not of their particular entries.

The paper is organized as follows. In the following Section \ref{sec:BD} we present the pertinent definitions and assumptions
that characterize the present approach to the heat conduction problem. The main result is obtained in Section \ref{sec:DF}
where we formulate the Jarzynski-type equation suited for the present problem and apply Jensen's inequality to derive two Clausius inequalities.
These inequalities immediately imply that, in the steady state and in the statistical mean, heat and entropy always flow from
the hotter to the colder bath. Section \ref{sec:EX} is devoted to an analytical example where a three-level system is alternately coupled to
two harmonic oscillators with different temperatures and the heat and entropy flows mentioned above can be explicitly calculated
by computer-algebraic means.
Moreover, in this example the steady state $p^{(s)}$ is characterized by two parameters $\beta^{(s)}$ and $\gamma^{(s)}$,
$\beta^{(s)}$ being some sort of inverse temperature and $\gamma^{(s)}$ characterizing the deviation of the steady state from a Gibbs state,
and the dependence of $\beta^{(s)}$ and $\gamma^{(s)}$ on the inverse temperatures of the two heat baths can be explicitly determined.
In the linear regime $\beta_0\approx \beta_1$, considered in Section \ref{sec:LR}, the steady state $p^{(s)}$ will only infinitesimally deviate
from the Gibbs state of inverse temperature $\beta_0$. This deviation can be calculated by means of first order perturbation theory
and used to determine the common slope $a$ of the functions of entropy transfer and (reduced) heat transfer at $\beta_1=\beta_0$.
The result, of course, depends of the eigenvalues and eigenvectors of $T^{(0)}$ which can be obtained for the analytical example
of Section \ref{sec:EX} but probably not, or only approximately, for ``real" systems.
We close with a summary and outlook in Section \ref{sec:SO}.

%%%%%%%%%%%%%%%%%%%%%%%%%%%%%%%%%%%%%%%%%%%%%%%%%%%%%%%%%%%%%%%%%%%%%%%%%%%%%%%%%%%%%%%%%%%%%%%%%%%%%%%%%%%%%%%%%%%%%%%%%%%%%%%%%%%%%%%%%%
\section{Basic definitions}\label{sec:BD}
%%%%%%%%%%%%%%%%%%%%%%%%%%%%%%%%%%%%%%%%%%%%%%%%%%%%%%%%%%%%%%%%%%%%%%%%%%%%%%%%%%%%%%%%%%%%%%%%%%%%%%%%%%%%%%%%%%%%%%%%%%%%%%%%%%%%%%%%%%

We consider an $N$-level system (NLS) described by a finite index set ${\mathcal N}$,
energies $E_n$ and degeneracies $d_n$ for $n\in{\mathcal N}$. The Hamiltonian for the NLS
without interaction with the heat baths will hence be of the form
\begin{equation}\label{Ham}
 H=\sum_n E_n\,{\sf P}_n
 \;,
\end{equation}
with a complete family of mutually orthogonal projections ${\sf P}_n,\;n\in{\mathcal N},$ and $d_n=\mbox{Tr }{\sf P}_n$.

The Gibbs states for the NLS with inverse temperature $\beta>0$ are characterized by probabilities
\begin{equation}\label{defpi}
  p_n^{(\beta)} = \frac{d_n}{Z^{(\beta)}}\,\exp\left(-\beta\,E_n \right)
  \;,
\end{equation}
where the partition function $Z^{(\beta)}$ is given by
\begin{equation}\label{defZ}
  Z^{(\beta)} =\sum_n d_n\,\exp\left(-\beta\,E_n \right)
  \;.
\end{equation}
We assume a rather general initial state $\rho^{(i)}$ of the NLS of the form
\begin{equation}\label{initialstate}
 \rho^{(i)}= \sum_n p_n^{(i)} \frac{{\sf P}_n}{d_n}
 \;,
\end{equation}
that is not necessarily a Gibbs state. After the interaction with the first heat bath we perform a L\"uders
measurement of the energy that, without selection according to the outcome, is assumed to yield the intermediate mixed state
\begin{equation}\label{firstmeas}
 \rho'=\sum_n q_n^{(i)} \frac{{\sf P}_n}{d_n}
 \;,
\end{equation}
such that the transition between the probability distributions $p^{(i)}\mapsto q^{(i)}$
is given by a left stochastic matrix $T^{(0)}$ characterizing the first heat bath according to
\begin{equation}\label{stochmat0}
  q^{(i)}_n = \sum_m T^{(0)}_{nm} \, p^{(i)}_m,\quad \mbox{ for all } n\in {\mathcal N}
  \;.
\end{equation}
Thereafter, an interaction with a second heat bath is assumed that, after a second L\"uders measurement of the energy,
results in the final state
\begin{equation}\label{finalstate}
 \rho^{(f)}= \sum_n p_n^{(f)} \frac{{\sf P}_n}{d_n}
 \;,
\end{equation}
such that, analogously to the first step,
\begin{equation}\label{stochmat1}
   p^{(f)}_n = \sum_m T^{(1)}_{nm} \, q^{(i)}_m,\quad \mbox{ for all } n\in {\mathcal N}
  \;,
\end{equation}
where $T^{(1)}$ is another (left) stochastic matrix characterizing the second heat bath.

If we model the heat baths by quantum systems and their successive interaction with the NLS by some unitary time evolution
the form (\ref{firstmeas}, \ref{finalstate}) of the intermediate state and the final state of the NLS does not follow
automatically but represents a crucial assumption of our approach.
Only if the NLS is non-degenerate, i.~e., all $d_n=1$, then (\ref{firstmeas}) and (\ref{finalstate})
will follow from general principles. In other cases, these assumptions may fail or be only approximately satisfied.

The stochastic matrices $T^{(0)}$ and $T^{(1)}$ generally possess invariant probability distributions or \textit{fixed points}, i.~e.,
eigenvectors corresponding to the eigenvalue $1$ with non-negative entries. At the moment we need not assume that these
probability distributions are unique, but we will rather postulate that there exist invariant probability distributions
of Gibbs type, i.~e.,
\begin{equation}\label{Gibbsinvariant}
  \sum_m T^{(i)}_{nm}   \, p_m^{(\beta_i)}=p_n^{(\beta_i)},\quad \mbox{ for all } n\in{\mathcal N} \mbox { and  } i=0,1
  \;.
\end{equation}
It will be plausible to identify the inverse temperatures $\beta_0$ and $\beta_1$ with the respective inverse temperatures
of the heat baths: According to the zeroth law of thermodynamics the interaction between a heat bath and an NLS of the same temperature
should not change the occupation probabilities of the energy levels.

The total transition $p^{(i)} \mapsto q^{(i)} \mapsto p^{(f)}$ will be achieved  by the  product matrix
\begin{equation}\label{defproduct}
  T = T^{(1)} \, T^{(0)}
  \;,
\end{equation}
which is again a stochastic matrix.
In the case of $\beta_0=\beta_1$ the Gibbs state probabilities $p^{(\beta_0)}_n$ would form an invariant probability
distribution of  $T$, but in the general case of $\beta_0 \neq \beta_1$ the fixed point of $T$ (not necessarily unique)
will be different from $p^{(\beta_i)},\;i=0,1$. Starting with an arbitrary initial distribution $p^{(i)}$ and iterating
the above-described protocol of successive interactions with the heat baths and energy measurements one would
obtain a sequence $T^\nu p^{(i)}$ of probability distributions that converge towards a fixed point of $T$:
\begin{equation}\label{fixedT}
  p^{(s)} =\lim_{\nu\to\infty }T^\nu p^{(i)}
  \;.
\end{equation}
If the fixed point is not unique, $ p^{(s)} $ will depend on the initial distribution $ p^{(i)}$. Physically,
$p^{(s)}$ represents a \textit{steady state} that will be asymptotically assumed. The NLS transports heat and entropy between
the two heat baths, that are approximately unchanged (otherwise the stochastic matrix $T$ could not be kept constant),
see Figure \ref{FIGTHB}.
This is similar as for the Carnot cycle of classical thermodynamics. The difference, however, is that
we do not consider additional extraction of work and allow for irreversible processes, whereas the Carnot process is reversible.

The steady state distribution $p^{(s)}$  can be used to calculate the steady flow of heat and entropy between the two heat baths
due to the process described above. A natural question is whether this flow is always directed from the hotter bath to the colder one,
as one would expect in accordance with the second law of thermodynamics. This question will be addressed in the next Section \ref{sec:DF}.

%%%%%%%%%%%%%%%%%%%%%%%%%%%%%%%%%%%%%%%%%%%%%%%%%%%%%%%%%%%%%%%%%%%%%%%%%%%%%%%%%%%%%%%%%%%%%%%%%%%%%%%%%%%%%%%%%%%%%%%%%%%%%%%%%%%%%%%%%%
\section{Direction of heat and entropy flow}\label{sec:DF}
%%%%%%%%%%%%%%%%%%%%%%%%%%%%%%%%%%%%%%%%%%%%%%%%%%%%%%%%%%%%%%%%%%%%%%%%%%%%%%%%%%%%%%%%%%%%%%%%%%%%%%%%%%%%%%%%%%%%%%%%%%%%%%%%%%%%%%%%%%
We recall the general Jarzynski-type equation that holds for transitions between $N$ states described by a stochastic matrix ${\sf T}$,
and interspersed with sequential measurements \cite{SSG21}:
\begin{equation}\label{Jar}
  \left\langle \frac{{\sf p}^{(0)}_n\,{\sf q}_m}{{\sf p}^{(0)}_m\,{\sf p}_n}\right\rangle =1
  \;.
\end{equation}
Here the terms in the bracket represent four random variables
\begin{equation}\label{randvar}
  Y_\nu: {\mathcal N}\times  {\mathcal N} \rightarrow {\mathbbm R}
  \;,
\end{equation}
defined on an ``elementary event space" $ {\mathcal N}\times  {\mathcal N} $ with probability function
\begin{eqnarray}\label{probfunc1}
 P&:&  {\mathcal N}\times  {\mathcal N} \rightarrow [0,1],\\
 \label{probfunc2}
 P(m,n)&:=& {\sf T}_{mn} {\sf p}_n
 \;,
\end{eqnarray}
where ${\sf p}$ is any positive probability distribution and $\left\langle \ldots \right\rangle$ denotes the expectation value
calculated by means of the $ P(m,n)$.
The four random variables occurring in the Jarzynski-type equation (\ref{Jar}) are
\begin{eqnarray}
\label{Y1}
  Y_1(m,n) &=& {\sf p}_n, \\
  \label{Y2}
  Y_2(m,n) &=&{\sf q}_m= \sum_n {\sf T}_{mn}\,{\sf p}_n, \\
  \label{Y3}
  Y_3(m,n) &=& {\sf p}_n^{(0)}, \\
  \label{Y4}
  Y_4(m,n) &=& {\sf p}_m^{(0)},
\end{eqnarray}
and ${\sf p}^{(0)}$ is a fixed point of ${\sf T}$, i.~e.,
\begin{equation}\label{fixT}
  \sum_n {\sf T}_{mn}\,{\sf p}^{(0)}_n={\sf p}^{(0)}_m
  \;,
\end{equation}
with positive components.

We will apply (\ref{Jar}) in two ways: The first one is given by
\begin{equation}\label{firstway}
  {\sf T}= T^{(0)},\; {\sf p}_n= p_n^{(s)},\; \mbox{ and } {\sf p}_n^{(0)}= p_n^{(\beta_0)}
  \;,
\end{equation}
such that (\ref{fixT}) holds according to (\ref{Gibbsinvariant}). Using Jensens's inequality (JI) and the fact that
$x\mapsto - \log x$ is a convex function, we conclude, analogously to \cite[(63 - 66)]{SSG21},
\begin{eqnarray}
  0 &=& -\log 1\stackrel{(\ref{Jar})}{=} -\log\left\langle \frac{p_n^{(\beta_0)}\,q_m^ {(s)}}{p_m^{(\beta_0)}\,p_n^{(s)}}\right\rangle\\
  &\stackrel{(JI)}{\le}& \left\langle
  -\log\frac{p_n^{(\beta_0)}}{d_n}+\log\frac{p_m^{(\beta_0)}}{d_m}-\log \frac{q_m^{(s)}}{d_m}+\log \frac{p_n^{(s)}}{d_n}
  \right\rangle\\
  &=& \left\langle \beta_0 E_n +\log Z^{(\beta_0)}-\beta_0 E_m-\log Z^{(\beta_0)}
  \right\rangle -\sum_m q_m^{(s)}\, \log\frac{q_m^{(s)}}{d_m}+\sum_n p_n^{(s)}\, \log\frac{p_n^{(s)}}{d_n}\\
  &=& \left\langle \Delta S  \right\rangle -\beta_0\left\langle \Delta Q \right\rangle
  \;.
  \end{eqnarray}
Here we have used the obvious definitions of the random variables ``heat transfer"
\begin{equation}\label{DeltaQranvar}
\Delta Q(m,n):= E_m-E_n
\;,
\end{equation}
and   ``entropy transfer"
\begin{equation}\label{DeltaSranvar}
\Delta S(m,n):= \log \frac{{\sf p}_n}{d_n}-\log \frac{{\sf q}_m}{d_m}
\;.
\end{equation}
Hence we have shown the following {\textit{Clausius inequality}:
\begin{equation}\label{Clausius1}
 \left\langle \Delta S  \right\rangle \ge \beta_0\left\langle \Delta Q \right\rangle
 \;.
\end{equation}
Note that the choice ${\sf p}_n= p_n^{(s)}$ is not necessary to derive (\ref{Clausius1}); similarly as in \cite{SSG21}
this Clausius inequality would hold for arbitrary initial probability distributions.
However, the choice of a steady state distribution will be crucial
to derive the second Clausius inequality, see (\ref{Clausius}) below.

The second way of applying (\ref{Jar})  will be given by the choice
 \begin{equation}\label{secondway}
  {\sf T}= T^{(1)},\; {\sf p}_n= \sum_m T^{(0)}_{nm}\,p_m^{(s)},\; \mbox{ and } {\sf p}_n^{(0)}= p_n^{(\beta_1)}
  \;,
\end{equation}
such that, again, (\ref{fixT}) holds according to (\ref{Gibbsinvariant}).
Due to the fixed point property of $p^{(s)}$ with respect to $T=T^{(1)}\,T^{(0)}$ the choice (\ref{secondway})
implies that the role of ${\sf p}_n$ and ${\sf q}_m$ is interchanged. This means that
at steady state the NLS absorbs the same amount of heat during contact with the first heat bath as it emits
during the contact with the second one, analogously for the entropy.  It follows that (\ref{Clausius1})
again holds, but with the replacements
\begin{eqnarray}
  \left\langle \Delta S  \right\rangle &\mapsto&-\left\langle \Delta S  \right\rangle, \\
  \left\langle \Delta Q  \right\rangle &\mapsto&-\left\langle \Delta Q  \right\rangle, \quad\mbox{ and}\\
  \beta_0 &\mapsto& \beta_1
  \;.
\end{eqnarray}
This gives
\begin{eqnarray}\label{Clausius2a}
 -\left\langle \Delta S  \right\rangle &\ge& -\beta_1 \left\langle \Delta Q \right\rangle\;,\\
 \label{Clausius2b}
 \left\langle \Delta S  \right\rangle &\le& \beta_1 \left\langle \Delta Q \right\rangle
 \;,
\end{eqnarray}
and thus, together with (\ref{Clausius1}), we obtain the \textit{Clausius inequalities}
\begin{equation}\label{Clausius}
 \beta_0 \left\langle \Delta Q \right\rangle \le \left\langle \Delta S  \right\rangle \le \beta_1 \left\langle \Delta Q \right\rangle
 \;.
\end{equation}
These inequalities have the same form as the corresponding inequalities derived in \cite{SSG21} and, under different assumptions,
in \cite{JW04,JR10,Sa20,S78,AMH21,SW21}, but generally a different meaning: For example,
in  \cite{SSG21} the inverse temperature $\beta_1$ refers to the NLS before the interaction
with the heat bath of inverse temperature $\beta_0$, but in (\ref{Clausius}) the inverse temperature $\beta_1$ refers to the second heat bath,
whereas the NLS is in the steady state and, in general, has no temperature at all.

Immediate consequences of (\ref{Clausius}) are
\begin{equation}\label{Clausius3}
 \left( \beta_1-\beta_0\right) \left\langle \Delta Q \right\rangle \ge 0
 \;,
\end{equation}
and
\begin{eqnarray}
\label{Clausius4a}
  \left( \beta_1-\beta_0\right) \left\langle \Delta S \right\rangle   &=&
  \beta_1\,\left\langle \Delta S \right\rangle-\beta_0\,\left\langle \Delta S \right\rangle \\
  \label{Clausius4b}
  &\stackrel{(\ref{Clausius})}{\ge}&  \beta_1\left(\beta_0 \left\langle \Delta Q \right\rangle \right)
  - \beta_0\left(\beta_1 \left\langle \Delta Q \right\rangle \right)
  \\
  \label{Clausius4c}
   &=& 0
   \;.
\end{eqnarray}
These equations imply that, in the steady state regime and in the statistical mean,
heat and entropy always flow from the hotter bath to the colder one,

\begin{figure}[htp]
\centering
\includegraphics[width=0.8\linewidth]{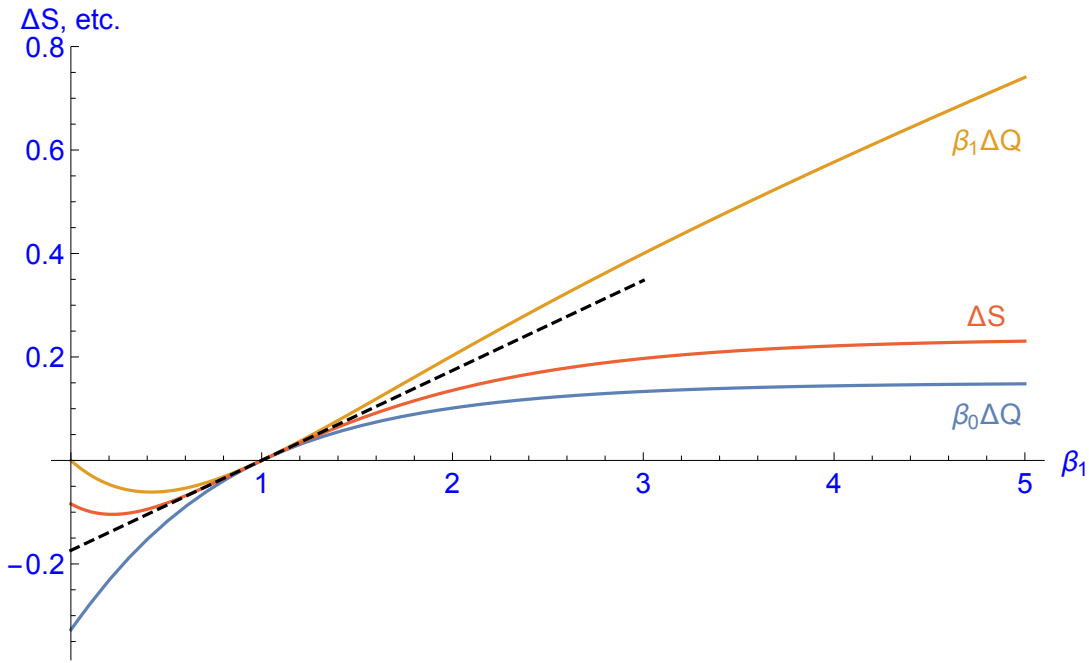}
\caption{Plot of the transferred entropy  $\left\langle\Delta S\right\rangle$ per cycle (red curve) and reduced heat transfer
$\beta_0\,\left\langle\Delta Q\right\rangle$ (blue curve) and $\beta_1\,\left\langle\Delta Q\right\rangle$
(dark yellow curve) as a function of the inverse temperature $\beta_1$ of the second
heat bath and fixed inverse temperature $\beta_0=1$ of the first one.
The common tangent at $\beta_1=\beta_0=1$ is indicated by the dashed black line with slope $a=0.173895$
that will be calculated in Section \ref{sec:LR}.
The calculations have been made for the analytical example of Section \ref{sec:EX}
and steady state regime. Note that the Clausius inequalities (\ref{Clausius}) are satisfied
and that the signs of $\left\langle\Delta S\right\rangle$ and   $\left\langle\Delta Q\right\rangle$ coincide
with the sign of $\beta_1-\beta_0$ in accordance with the second law of thermodynamics.}
\label{FIGDSQ}
\end{figure}

%%%%%%%%%%%%%%%%%%%%%%%%%%%%%%%%%%%%%%%%%%%%%%%%%%%%%%%%%%%%%%%%%%%%%%%%%%%%%%%%%%%%%%%%%%%%%%%%%%%%%%%%%%%%%%%%%%%%%%%%%%%%%%%%%%%%%%%%%%
\section{Analytical example}\label{sec:EX}
%%%%%%%%%%%%%%%%%%%%%%%%%%%%%%%%%%%%%%%%%%%%%%%%%%%%%%%%%%%%%%%%%%%%%%%%%%%%%%%%%%%%%%%%%%%%%%%%%%%%%%%%%%%%%%%%%%%%%%%%%%%%%%%%%%%%%%%%%%

As an example where the time average of the transition matrix ${\sf T}(\beta)$ corresponding
to a heat bath of inverse temperature $\beta$ can be exactly calculated
we have considered in \cite{SSG21} a single spin with spin quantum number $s=1$
coupled to a harmonic oscillator that serves as a heat bath. Hence we have a $N=3$-level system with ${\mathcal N}=\{1,0,-1\}$.
The total Hamiltonian and further details are given in \cite{SSG21} and need not be repeated here.
The Hamiltonian  strongly resembles the Jaynes-Cummings model \cite{JC63,GE90}
originally describing the interaction of a $2$-level system with a quantized radiation field
but also considered for $3$-level systems  \cite{AB95,TLV15}.
We reproduce the analytical result \cite{SSG21} for the entries of ${\sf T}(\beta)$:
 \begin{eqnarray}
\label{T11}
   {\sf T}_{1,1}(\beta) &=& \frac{1}{32} \left(e^{\beta}-1\right) \left(\frac{12}{e^{\beta}-1}+8
   e^{\frac{\beta}{2}} \coth ^{-1}\left(e^{\frac{\beta}{2}}\right)+3 e^{-\beta} \Phi \left(e^{-\beta},2,\frac{3}{2}\right)-8\right), \\
   \label{T12}
   {\sf T}_{1,0}(\beta) &=& \frac{1}{4} \left(1-2 \sinh \left(\frac{\beta}{2}\right) \tanh
   ^{-1}\left(e^{-\frac{\beta} {2}}\right)\right), \\
   \label{T13}
  {\sf T}_{1,-1}(\beta)&=& \frac{3}{32} e^{-3 \beta0} \left(4 e^{\beta }-\left(e^{\beta }-1\right) \Phi
   \left(e^{-\beta},2,\frac{3}{2}\right)\right),\\
   \label{T21}
  {\sf T}_{0,1}(\beta)&=&\frac{1}{4} e^{\beta} \left(1-2 \sinh \left(\frac{\beta }{2}\right) \tanh
   ^{-1}\left(e^{-\frac{\beta}{2}}\right)\right),\\
   \label{T22}
   {\sf T}_{0,0}(\beta)&=& \frac{1}{2},\\
   \label{T23}
   {\sf T}_{0,-1}(\beta)&=&\frac{1}{4} e^{-\frac{3 \beta}{2}} \left(e^{\frac{\beta}{2}}+
   \left(e^{\beta}-1\right) \tanh ^{-1}\left(e^{-\frac{\beta}{2}}\right)\right),\\
   \label{T31}
  {\sf T}_{-1,1}(\beta)&=&\frac{3}{32} e^{-\beta} \left(4 e^{\beta}-\left(e^{\beta}-1\right) \Phi
   \left(e^{-\beta },2,\frac{3}{2}\right)\right),\\
   \label{T32}
  {\sf T}_{-1,0}(\beta)&=&\frac{1}{4} \left(2 \sinh \left(\frac{\beta }{2}\right) \tanh
   ^{-1}\left(e^{-\frac{\beta}{2}}\right)+1\right),\\
   \nonumber
  {\sf T}_{-1,-1}(\beta)&=&\frac{1}{32} e^{-3 \beta } \left(4 e^{2 \beta} \left(11 \sinh \left(\beta
   _0\right)+5 \cosh \left(\beta \right)-4 \sinh \left(\frac{\beta }{2}\right)
   \coth ^{-1}\left(e^{\frac{\beta }{2}}\right)-2\right)+3 \left(e^{\beta}-1\right) \Phi \left(e^{-\beta },2,\frac{3}{2}\right)\right)
   \;.\\
   \label{T33}&&
 \end{eqnarray}
 Here $\Phi(z,s,a):= \sum_{k\in{\mathbbm N}} \frac{z^k}{(k+a)^s}$  denotes Lerch's transcendent, see
\cite[$\S\, 25.14$]{NIST}.
It can be shown that ${\sf T}(\beta)$ is a left stochastic matrix and leaves the probability distribution
\begin{equation}\label{Gibbsbeta}
 p^{(\beta)}=\frac{1}{e^{-\beta}+1+e^{\beta}}\left(e^{-\beta},1,e^{\beta }\right)
\end{equation}
 invariant that corresponds to a Gibbs state with inverse temperature $\beta$.

\begin{figure}[htp]
\centering
\includegraphics[width=0.8\linewidth]{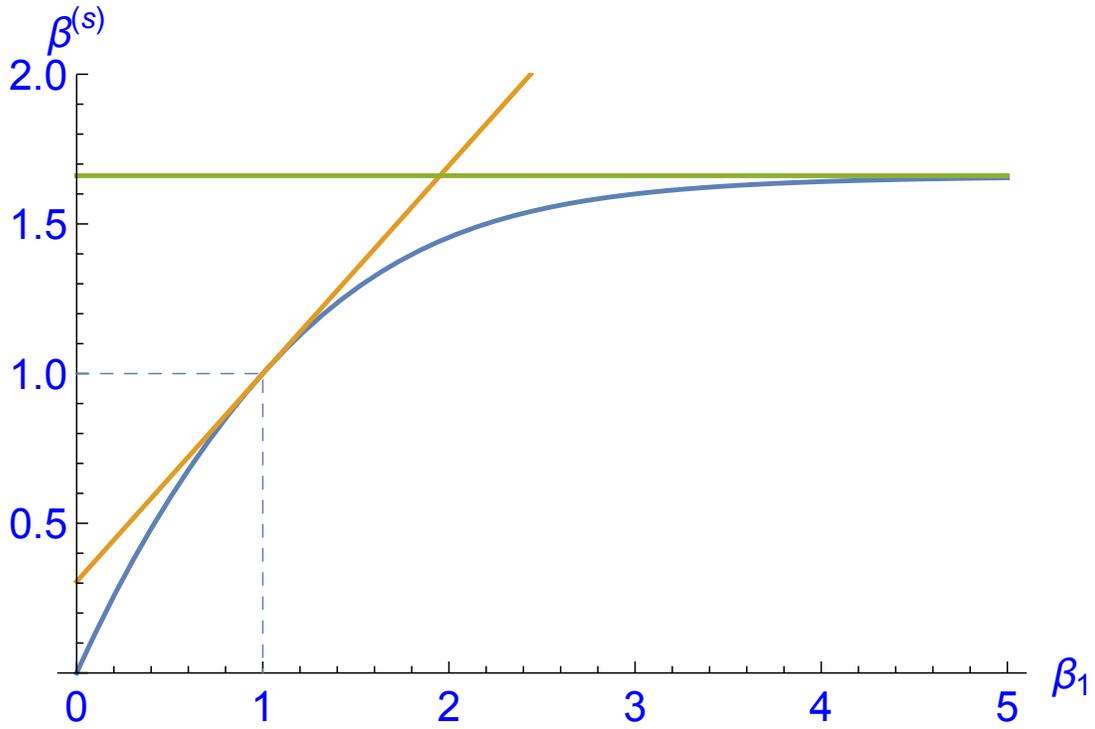}
\caption{Plot of the first parameter $\beta^{(s)}$ characterizing the steady state distribution $p^{(s)}$ of the analytical example as a function
of $\beta_1$ for fixed $\beta_0=1$ (blue curve). The slope $a_1=0.694846\ldots$ of the tangent at $\beta_1=\beta_0$ (dark yellow line) and
the asymptotic value of $\lim_{\beta_1\to\infty}\beta^{(s)}=1.66098\ldots$ (green line) have been calculated analytically.
}
\label{FIGBS}
\end{figure}

\begin{figure}[htp]
\centering
\includegraphics[width=0.8\linewidth]{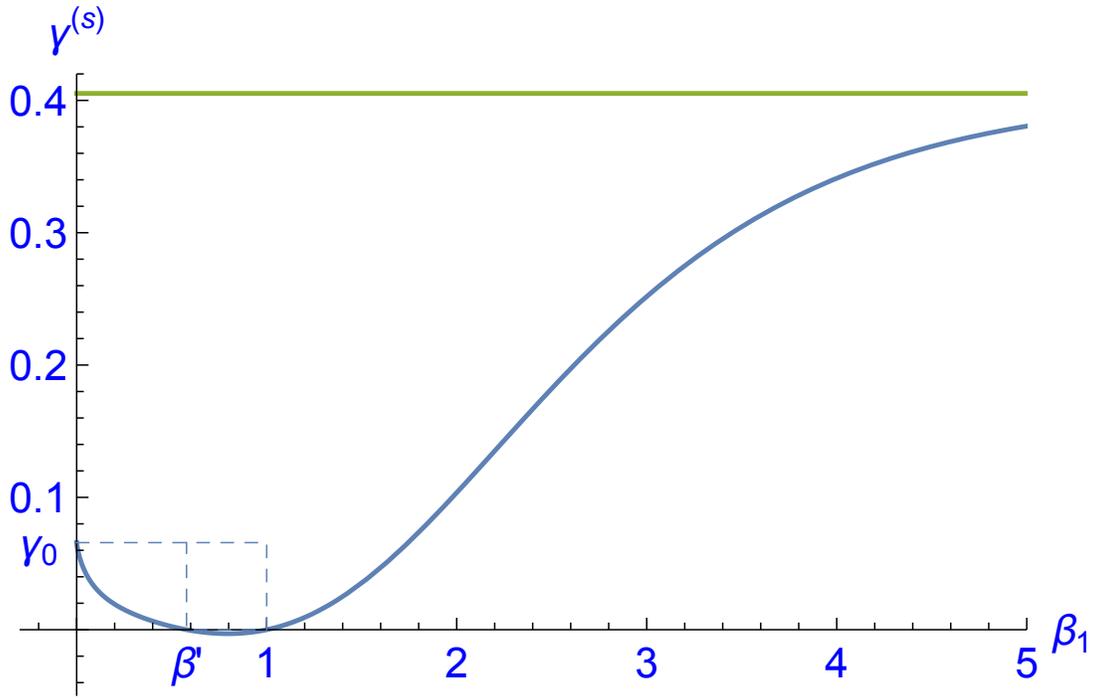}
\caption{Plot of the second parameter $\gamma^{(s)}$ characterizing the steady state distribution $p^{(s)}$ of the analytical example as a function
of $\beta_1$ for fixed $\beta_0=1$ (blue curve). The value of $\gamma^{(s)}=\gamma_0=0.0658087\ldots$ for $\beta_1=0$ and
the asymptotic value of $\lim_{\beta_1\to\infty}\gamma^{(s)}=0.405262\ldots$ (green line) have been calculated analytically.
For $\beta_1=\beta_0=1$ the parameter $\gamma^{(s)}$ vanishes since the corresponding Gibbs-type probability distribution is a fixed point of $T$.
Interestingly, there exists a second zero of $\gamma^{(s)}$ at $\beta_1=\beta'=0.578258\ldots$. In the interval $\beta' < \beta_1 <1$ the
parameter $\gamma^{(s)}$ has  small negative values and hence the function $E_n\mapsto \log p_n^{(s)}$ is slightly concave.
}
\label{FIGGS}
\end{figure}

Using this analytical result we set $T^{(0)}={\sf T}(\beta_0)$ and $T^{(1)}={\sf T}(\beta_1)$ for the model with
two heat baths and can explicitly calculate the heat and entropy transfer $\left\langle \Delta S\right\rangle (\beta_1)$
and $\beta_i\,\left\langle \Delta Q\right\rangle (\beta_1)$, considered in  Section \ref{sec:DF}, as functions of $\beta_1$.
Although the result is too complex to be shown here these transfer functions can be plotted for fixed values of $\beta_0$, say, $\beta_0=1$,
see Figure \ref{FIGDSQ}. Thus one can graphically confirm the Clausius inequalities (\ref{Clausius}) and the correct sign
of heat and entropy flow. The three functions have a common tangent at $\beta_0=\beta_1$ the slope of which will be calculated in Section
\ref{sec:LR}.

In the limit $\beta\to 0$ we obtain the bi-stochastic matrix
\begin{equation}\label{Tlimitzero}
 \lim_{\beta\to 0}{\sf T}(\beta) =
 \frac{1}{8}\left(
\begin{array}{ccc}
 3 & 2 & 3 \\
 2 & 4 & 2 \\
 3 & 2 & 3 \\
\end{array}
\right)
\;,
\end{equation}
whereas for $\beta\to\infty$
\begin{equation}\label{Tlimitinfty}
 \lim_{\beta\to\infty}{\sf T}(\beta) =
 \frac{1}{6}
 \left(
\begin{array}{ccc}
 3 & 0 & 0 \\
 1 & 3 & 0 \\
 2 & 3 & 6 \\
\end{array}
\right)
\;.
\end{equation}
Both limits facilitate the analytical calculation of corresponding asymptotic values of the pertinent functions.
By the way, the latter one (\ref{Tlimitinfty})
is an example of a stochastic matrix where the algebraic and geometric degeneracy of the eigenvalues no longer coincide:
the eigenvalue $t_2=1/2$ is doubly degenerate but possesses only a one-dimensional eigenspace ${\mathbbm R}\,(0,1,-1)^\top$.

Next we will investigate the structure of the steady state distribution $p^{(s)}$ as a function of $\beta_0$ and $\beta_1$
for the present example. A general probability distribution $p$ of a $3$-level system with energies $E_1,E_0,E_{-1}$
can be characterized by two parameters $\beta, \gamma$ such that
\begin{equation}\label{probpar}
 p_n = \frac{1}{Z} \exp\left(-\beta E_n +\gamma E_n^2 \right),\quad  \mbox{ for } n=1,0,-1
 \;,
\end{equation}
where $Z$ is chosen such that $p_1+p_0+p_{-1}=1$. A Gibbs-type probability distribution would thus be characterized by
$\gamma=0$ and $\beta$ being the inverse temperature of the distribution.
It is then a straightforward task to calculate the parameters $\beta^{(s)}, \gamma^{(s)}$ corresponding
to the steady state distribution $p^{(s)}$ as a function of $\beta_0$ and $\beta_1$. Again, the result is too complex to be shown,
but it can be plotted for fixed $\beta_0$, say, $\beta_0=1$, see Figures \ref{FIGBS} and \ref{FIGGS}.
Surprisingly, we find that for $\beta_0=1$ and $\beta_1=\beta'=0.578258\ldots$ the second parameter $\gamma^{(s)}$ vanishes
and hence the corresponding steady state distribution is of Gibbs type.

%%%%%%%%%%%%%%%%%%%%%%%%%%%%%%%%%%%%%%%%%%%%%%%%%%%%%%%%%%%%%%%%%%%%%%%%%%%%%%%%%%%%%%%%%%%%%%%%%%%%%%%%%%%%%%%%%%%%%%%%%%%%%%%%%%%%%%%%%%
\section{The linear regime}\label{sec:LR}
%%%%%%%%%%%%%%%%%%%%%%%%%%%%%%%%%%%%%%%%%%%%%%%%%%%%%%%%%%%%%%%%%%%%%%%%%%%%%%%%%%%%%%%%%%%%%%%%%%%%%%%%%%%%%%%%%%%%%%%%%%%%%%%%%%%%%%%%%%
So far, we have not made any assumptions about the relationship of the stochastic matrices $T^{(0)}$ and $T^{(1)}$.
They can be completely independent except that they have Gibbs-type fixed points.
If $T^{(0)}$ and $T^{(1)}$ were derived from a physical model of the interaction of the NLS with a heat bath
of inverse temperature $\beta$, as in the example of Section \ref{sec:EX},
there would probably be further relations between  $T^{(0)}$ and $T^{(1)}$, but we have not needed to consider them so far.
However, if we turn to the linear regime where the value of $\delta:=\beta_1-\beta_0$ is infinitesimally small,
we will need the following mild assumption, that is fortunately satisfied for the analytical example of Section \ref{sec:EX}:
\begin{assu}\label{ASS1}
 There exists a smooth family $\delta\mapsto \mathbf{T}(\delta)$ of (left) stochastic matrices
 defined for all $\left|\delta\right|<r$  and some $r>0$  such that
 \begin{equation}\label{family}
   T^{(0)}=\mathbf{T}(0),\quad\mbox{ and}\quad T^{(1)}=\mathbf{T}(\delta)=\mathbf{T}(0)+\delta \dot{\mathbf{T}}(0)+O(\delta^2)
   \;.
 \end{equation}
\end{assu}
Here the overdot $\dot{}$ means $\frac{\partial}{\partial \delta}\left(\ldots\right)\left.\right|_{\delta=0}$.
We will henceforward omit the argument $\delta=0$ for sake of simplicity. For example, for the total cycle we have
\begin{equation}\label{Tlinreg}
 T= T^{(1)}\,T^{(0)}=\mathbf{T}(\delta)\,\mathbf{T}= \mathbf{T}^2+ \delta\, \dot{\mathbf{T}}\, {\mathbf{T}}+O(\delta^2)
 \;.
\end{equation}
The steady state distribution $p^{(s)}$ will also be chosen as a smooth family denoted by
${\mathbf p}(\delta)$ such that $T\,p^{(s)}=p^{(s)}$ now reads
\begin{equation}\label{steadyp}
 \mathbf{T}(\delta)\,\mathbf{T}\,{\mathbf p}(\delta)= {\mathbf p}(\delta)
 \;,
\end{equation}
and
\begin{equation}\label{pdelta}
 {\mathbf p}(\delta) = {\mathbf p} + \delta\, \dot{{\mathbf p}} +O(\delta^2)
 \;,
\end{equation}
where ${\mathbf p}=p^{(\beta_0)}$.

We will calculate  $\dot{{\mathbf p}}$ by means of first order perturbation theory.
This is similar in concept to the derivation of a Kubo formula in Liouville space in \cite{MGM04},
where the temperature gradient introduced by the two heat baths is treated as an external perturbation.
Since the unperturbed matrix will, in general, not be symmetric, some slight modifications of the
Rayleigh-Schr\"odinger perturbation theory known from quantum mechanics are required that will be derived in what follows.
To simplify these calculations
we will restrict ourselves to the ``generic case" of
\begin{assu}\label{ASS2}
The eigenvalue $t_1=1$ of ${\mathbf T}$ is algebraically non-degenerate and for the other eigenvalues $t_2,\ldots$ the algebraic and geometric
degeneracy coincides.
\end{assu}
Hence there exist a complete basis of ${\mathbbm R}^N$ of (right) eigenvectors $\left| {\mathbf e}_n\right\rangle$ of ${\mathbf T}$
satisfying
\begin{equation}\label{eigenT}
  {\mathbf T}\,\left| {\mathbf e}_n\right\rangle = t_n\, \left| {\mathbf e}_n\right\rangle
  \;,
\end{equation}
for $n=1,\ldots, N$, where $\left| {\mathbf e}_1\right\rangle={\mathbf p}=p^{(\beta_0)}$ and $t_1=1$.
Here we have switched to Dirac's bra-ket style which makes the notation somewhat easier.
The dual basis of left eigenvectors of ${\mathbf T}$ will be denoted by $\left\langle {\mathbf e}^n\right|,\,n=1,\ldots,N$,
such that the following holds:
\begin{equation}\label{innerproducts}
 \left\langle {\mathbf e}^n \left| \right. {\mathbf e}_m\right\rangle =\delta^n_m\quad\mbox{(Kronecker delta)}
 \;,
\end{equation}
and
\begin{equation}\label{resolution}
  {\mathbbm 1}=\sum_{n=1}^{N}\left| {\mathbf e}_n\right\rangle \left\langle {\mathbf e}^n\right|
  \;.
\end{equation}
The latter implies
\begin{equation}\label{Teigenrep}
{\mathbf T}={\mathbf T}\, {\mathbbm 1}=\sum_{n=1}^{N}{\mathbf T}\,\left| {\mathbf e}_n\right\rangle \left\langle {\mathbf e}^n\right|
\stackrel{(\ref{eigenT})}{=} \sum_{n=1}^{N}t_n\,\left| {\mathbf e}_n\right\rangle \left\langle {\mathbf e}^n\right|
\;.
\end{equation}
Obviously, $\left\langle {\mathbf e}^1\right|=(1,1,\ldots,1)$. Let $\left\langle {\mathbf e}^1\right|^\perp$ denote
the subspace
\begin{equation}\label{defeperp}
\left\langle {\mathbf e}^1\right|^\perp
:= \{{\mathbf x}\in{\mathbbm R}^N \left|\left\langle{\mathbf e}^1 |{\mathbf x} \right\rangle=\sum_{n=1}^{N}x_n=0 \right.\}
\;.
\end{equation}
Upon differentiating (\ref{steadyp}) we obtain
\begin{equation}\label{steadypdiff}
  \dot{\mathbf T}\,{\mathbf T}\,{\mathbf p}+{\mathbf T}^2\,\dot{\mathbf p} =\dot{\mathbf p}
  \;,
\end{equation}
and hence, using ${\mathbf T}\,{\mathbf p}={\mathbf p}$,
\begin{equation}\label{pdotdet}
  \left( {\mathbbm 1}-{\mathbf T}^2\right)\,\dot{\mathbf p} = \dot{\mathbf T}\,{\mathbf p}
  \;.
\end{equation}
The two vectors $\dot{\mathbf p}$ and $\dot{\mathbf T}\,{\mathbf p}$ lie in the subspace $\left\langle {\mathbf e}^1\right|^\perp$
and the latter is left invariant under the linear map $ \left( {\mathbbm 1}-{\mathbf T}^2\right)$.
Hence (\ref{pdotdet}) can be solved for $\dot{\mathbf p}$ by inverting the restriction of $ \left( {\mathbbm 1}-{\mathbf T}^2\right)$
to the subspace $\left\langle {\mathbf e}^1\right|^\perp$. By means of (\ref{Teigenrep}) the result can be written as
\begin{equation}\label{pdotexpl}
 \dot{\mathbf p}=\sum_{n=2}^{N}\frac{1}{1-t_n^2}\,\left| {\mathbf e}_n\right\rangle
 \langle {\mathbf e}^n | \dot{\mathbf T} | {\mathbf p} \rangle
 \;.
\end{equation}

This completes the calculation of the steady state distribution in linear order in $\delta=\beta_1-\beta_0$.
Next we consider the corresponding energy transfer $\left\langle \Delta Q \right\rangle$.
Let ${\mathbf E}\in{\mathbbm R}^N$ be the ``energy" vector with components $\left\langle n\left|{\mathbf E}\right.\right\rangle=E_n$.
The expectation value $Q_0$ of the energy of the NLS before the contact with the first heat bath is
\begin{equation}\label{Q0}
Q_0=\sum_n p_n(\delta)\, E_n = \sum_n p_n\,E_n + \delta\, \sum_n \dot{p}_n\,E_n +O(\delta^2)
\;.
\end{equation}
After the contact with the first heat bath and a subsequent energy measurement we obtain the energy expectation value
\begin{eqnarray}
\label{Q1a}
  Q_1 &=& \sum_n q_n\,E_n=\sum_{nm}T^{(0)}_{nm}\,p_m(\delta)\,E_n\\
  \label{Q1b}
   &=& \sum_{nm} {\mathbf T}_{nm} \,p_m\,E_n + \delta\, \sum_{nm} {\mathbf T}_{nm} \,\dot{p}_m\,E_n+O(\delta^2)\\
   \label{Q1c}
   &=& \sum_n p_n\,E_n + \delta\, \sum_{nm} {\mathbf T}_{nm} \,\dot{p}_m\,E_n+O(\delta^2)
   \;.
\end{eqnarray}
Hence
\begin{equation}\label{DQlin}
 \left\langle \Delta Q \right\rangle= Q_1-Q_0=:\delta\,a +O(\delta^2)
\end{equation}
where
\begin{eqnarray}
\label{DQ1a}
  a &=&\sum_{nm} {\mathbf T}_{nm} \,\dot{p}_m\,E_n-\sum_n \dot{p}_n\,E_n\\
  \label{DQ1b}
  &=& \left\langle {\mathbf E}\left| {\mathbf T}-{\mathbbm 1}\right| \dot{\mathbf p}\right\rangle\\
  \label{DQ1c}
  &\stackrel{(\ref{pdotexpl})}{=}&\langle {\mathbf E}\,
  |\sum_{n=2}^{N}\frac{t_n-1}{1-t_n^2}\, {\mathbf e}_n\,\rangle \langle {\mathbf e}^n | \dot{\mathbf T} | {\mathbf p} \rangle\\
  \label{DQ1d}
   &=& -\sum_{n=2}^{N}\frac{1}{1+t_n}\,\langle {\mathbf E}\left| \right.{\mathbf e}_n\,
   \rangle \langle {\mathbf e}^n | \dot{\mathbf T} | {\mathbf p} \rangle
   \;.
\end{eqnarray}
This completes the calculation of the slope $a$ of the tangent of $\beta_1\mapsto\left\langle \Delta Q \right\rangle$
at $\beta_1=\beta_0$ which can be considered as a sort of heat conduction coefficient.
The fact that $a\ge 0$ is a consequence of (\ref{Clausius3}).

This calculation establishes an ``external Fourier law", similarly as in \cite{MGM04}, that holds independent
of a possibly constant temperature gradient inside the NLS.

In the example of Section \ref{sec:EX} and for $\beta_0=1$ the right and left eigenvectors of ${\mathbf T}$ as well as $\dot{\mathbf T}$
can be explicitly calculated although not be shown here due to their overwhelming complexity. Inserting these terms into
(\ref{DQ1d}) gives $a=0.173895\ldots$, the slope of the common tangent of the three function plots in Figure \ref{FIGDSQ}.

%%%%%%%%%%%%%%%%%%%%%%%%%%%%%%%%%%%%%%%%%%%%%%%%%%%%%%%%%%%%%%%%%%%%%%%%%%%%%%%%%%%%%%%%%%%%%%%%%%%%%%%%%%%%%%%%%%%%%%%%%%%%%%%%%%%%%%%%%%
\section{Summary and outlook}\label{sec:SO}
%%%%%%%%%%%%%%%%%%%%%%%%%%%%%%%%%%%%%%%%%%%%%%%%%%%%%%%%%%%%%%%%%%%%%%%%%%%%%%%%%%%%%%%%%%%%%%%%%%%%%%%%%%%%%%%%%%%%%%%%%%%%%%%%%%%%%%%%%%
The second law of thermodynamics has several formulations of varying generality;
one of them is that heat always flows spontaneously from the hotter to the colder body.
In the field of quantum mechanics, the formulation, let alone the proof, of a corresponding second law is highly
obscure and controversial, despite (or even because of) the large body of literature on this subject.
In this situation, it seems advisable to gather a few facts that capture what we know for sure.
We know, that the von Neumann entropy increases in the course of a L\"uders measurement \cite{vN32}.
Further, again in the case of sequential  L\"uders measurements, there hold various Jarzynski-type equations  and
for each of them Jensen's inequality produces second law-like statements, see \cite{SG20a,SG20b}.
In this paper, we have singled out a particular Jarzynski-type equation so that the corresponding second law-like statement
essentially coincides with the above formulation, namely, that heat flows from the hotter to the colder heat bath.

However, although this result is very general and relatively easy to prove, we need to mention some restricting
assumptions that were required for the proof.
The quantum system is an arbitrary $N$-level system (NLS) and we implicitly assume that its states are
completely characterized by the occupation probabilities $p_n$ of the $n$-th energy level,
not only at the beginning but also during the whole heat conduction process.
The two heat baths are not analyzed in detail; they almost appear as ``black boxes" only described by stochastic matrices
that contain the conditional transition probabilities between the energy levels of the NLS and leave invariant
Gibbs states of inverse temperature $\beta_i, i=0,1$.
For technical reasons, we had to alternate contacts of the NLS with the two heat baths,
since we would lose the overview of the heat and entropy transfer with a total contact of all three systems.
Hopefully, this special protocol will be a reasonable approximation for real heat conduction processes.
A a benefit of our approach we mention that no thermalization assumptions are needed; the heat conduction process
can be arbitrarily irreversible. Our calculations are essentially confined to the steady state regime.
By the way, the approach to a steady state after a large number of cycles
is not an extra assumption but follows from the property that all eigenvalues of a stochastic matrix except the eigenvalue $1$
are $<1$ in absolute value.

It looks promising to apply the present approach to similar problems.
For example, a general thermodynamic process could be split into pure heat and work parts that can be represented by
(bi-)stochastic matrices analogously as in this paper and the Jarzynski-Jensen method could be used to obtain second law-like
statements.

\section*{Acknowledgment}
%%%%%%%%%%%%%%%%%%%%%%%%%%%%%%%%%%%%%%%%%%%%%%%%%%%%%%%%%%%%%%%%%%%%%%%%%%%%%%%%%%%%%%%%%%%%%%%%%%%%%%%%%%%%%%%%%%%%%%%%%%%%%%%%%%%%%%%%
This work was funded by the Deutsche Forschungsgemeinschaft (DFG) Grant No.~GE 1657/3-1.
We sincerely thank the members of the DFG Research Unit FOR2692 for fruitful discussions.
In addition, we are grateful to Philipp Strasberg for expert comments and literature references.

%%%%%%%%%%%%%%%%%%%%%%%%%%%%%%%%%%%%%%%%%%%%%%%%%%%%%%%%%%%%%%%%%%%%%%%%%%%%%%%%%%%%%%%%%%%%%%%%%%%%%%%%%%%%%%%%%%%%%%%%%%%%%%%%%%%%%%%%

\end{document}